\documentclass[prl,twocolumn,showpacs,superscriptaddress,amsmath,amssymb]{revtex4-2}
\usepackage[utf8]{inputenc}
\usepackage{amsthm}
\usepackage{qcircuit}
\usepackage{epsfig}
\usepackage{booktabs,multirow}
\usepackage{pdfcomment}

\newcommand{\bra}[1]{{\langle{#1}|}}
\newcommand{\ket}[1]{{|{#1}\rangle}}

\usepackage[caption=false]{subfig}
\usepackage{graphicx,bm}
\usepackage{verbatim}
\begin{document}
	\title{Selective continuous-variable quantum process tomography}
	\author{Virginia Feldman}
	\affiliation{Instituto de F\'isica, Facultad de Ingenier\'ia, Universidad de la Rep\'ublica, Montevideo, Uruguay}
	\author{Ariel Bendersky}
	\affiliation{Instituto de Investigaci\'on en Ciencias de la Computaci\'on (ICC), CONICET-UBA, Buenos Aires, Argentina}
 \affiliation{Departamento de Computaci\'on, Facultad de Ciencias Exactas y Naturales, Universidad de Buenos Aires, Buenos Aires, Argentina}
 \affiliation{Quantum Research Center, Technology Innovation Institute, Abu Dhabi, UAE}

	\begin{abstract}
		Quantum process tomography is a useful tool for characterizing quantum processes. This task is essential for the development of different areas, such as quantum information processing.  In this work, we present a protocol for selective continuous-variable quantum process tomography. Our proposal allows one to selectively estimate any element of an unknown continuous-variable quantum process in the position representation, without requiring the complete reconstruction of the process. By resorting to controlled squeezing and translation operations, and adaptatively discretizing the process, a direct measure of an estimate of any process element can be obtained. Furthermore, we show, supported by numerical simulations, how the protocol can be used to partially reconstruct on a region a continuous-variable quantum process.
		
	\end{abstract}

	\maketitle

\section{I. Introduction}
Quantum information processing, quantum computing and quantum error correction rely to a great extent on the correct characterization of quantum processes. The methods used for this characterization are commonly referred to as quantum process tomography (QPT) \cite{nielsen2002,chuang1997,poyatos1997,altepeter2003,mohseni2006,mohseni2008}.

QPT in infinite dimensions has primarily focused on reconstructing optical processes in the discrete Fock basis. The most widely known of these techniques makes use of a set of coherent states \cite{lobino2008complete,rahimi2011quantum,anis2012maximum,kumar2013experimental,lobino2009memory} or squeezed states \cite{fiuravsek} as probe states to achieve this reconstruction. 
%QPT with coherent state was employed to characterize attenuation and phase shift channels \cite{lobino2008complete}, photon creation and annihilation operations \cite{kumar2013experimental} and a quantum optical memory \cite{lobino2009memory}. 
Specifically, the method of coherent-state quantum process tomography was first introduced in \cite{lobino2008complete} and later adapted in \cite{rahimi2011quantum}. This approach is based on the notion that any density matrix can be expressed as a linear combination of coherent states through the Glauber-Sudarshan $P$ function \cite{glauber1963,sudarshan1963}. Therefore, by knowing how the process acts on every coherent state, the process output for any input state can be determined. These methods employ homodyne quantum state tomography combined with maximum-likelihood reconstruction to find the effect of the unknown process on a set of coherent states. The numerical reconstruction procedures involve determining the density matrices of the output states for each input state and integrating with the $P$ function. An adaptation of this protocol, which uses statistical inference to directly reconstruct the  process in the Fock basis from the experimental data, was proposed in \cite{anis2012maximum}.
All these proposals are non-selective, requiring the full reconstruction of the process for a subspace spanned by Fock states up to a certain cut-off value, correlated with the maximum amplitude of the set of coherent probe states. Working with a finite set of coherent states, thus with limited amplitude, can lead to truncation errors and an inaccurate reconstruction of the process. Additionally, the use of the $P$ function can make the computation of the reconstructed process difficult due to its highly singular nature \cite{kumar2013experimental,Damanet2018}.

To date, there are no protocols in the literature for selective quantum process tomography in the position representation.

Following a similar approach to that in  \cite{feldman2022selective}, which introduced an algorithm for selective quantum state tomography for continuous-variable systems, this article adopts a continuous-variable approach by working in the position representation. The action of a continuous-variable quantum process $\mathcal{E}$ on a state $\rho$ can be expressed as
\begin{equation}
 \mathcal{E}(\rho)=\int\mathcal{E}_{\substack{x,y\\w,z}}\ket{x}\bra{y}\rho\ket{w}\bra{z}dxdydwdz,
 \label{processdef1}
   \end{equation}
where $\mathcal{E}_{\substack{x,y\\w,z}}$ is a complex-valued function. We will refer to the values $\mathcal{E}_{\substack{x,y\\w,z}}$ as elements of the process matrix, drawing an analogy to the $\chi$-matrix description of a quantum process in the finite-dimensional case \cite{nielsen2002}. Therefore, $\mathcal{E}_{\substack{x,y\\w,z}}$ will also be referred to as elements of the quantum process. The knowledge of $\mathcal{E}_{\substack{x,y\\w,z}}$ provides a complete description of the process.  Taking this into consideration, we present a protocol to selectively estimate any element $\mathcal{E}_{\substack{x,y\\w,z}}$ of this description. The estimation can be obtained directly from measurements, without resorting to the use of a set of probing states and without going through the full reconstruction of the process.

The article is organized as follows. In Section II, our protocol for selective continuous-variable quantum process tomography is presented. In Section III, numerical simulations for the partial reconstruction on a region of different processes by application of the algorithm are shown. Finally, Section IV includes a discussion and concluding remarks.

\section{II. Selective continuous-variable quantum process tomography}

The algorithm we present allows one to selectively estimate any element $\mathcal{E}_{\substack{x,y\\w,z}}$ of the process $\mathcal{E}$ as defined in \eqref{processdef1}. The proposed quantum circuit for estimating an element $\mathcal{E}_{\substack{a,b\\c,d}}$ makes use of a control qubit and relies on controlled and anticontrolled translation $T$ and squeezing $S$ gates, as depicted in Fig. \ref{circuitCVproceso}.

	\begin{figure*}
 \centering
        \large
        \hspace{-0.5cm}
        \Qcircuit @C=0.4em @R=2.3em {
        \lstick{\ket{\psi}} & \qw \ar@{.}[]+<0.9em,1em>;[d]+<0.9em,-1em> & \gate{S(r_c)} & \gate{T(c)} & \gate{S(r_b)} & \gate{T(b)} & \qw \ar@{.}[]+<0em,1em>;[d]+<0em,-1em> &  \gate{\mathcal{E}} & \qw  &\gate{T^{\dagger}(d)} & \gate{S^{\dagger}(r_d)} & \gate{T^{\dagger}(a)} & \gate{S^{\dagger}(r_a)} &\qw  \ar@{.}[]+<-0.1em,1em>;[d]+<-0.1em,-1em> & \rstick{\Pi_{\delta}} \\
        \lstick{\ket{0}} & \gate{H} & \ctrl{-1} & \ctrl{-1} & \ctrlo{-1} & \ctrlo{-1} & \qw & \qw& \qw & \ctrl{-1} & \ctrl{-1}  & \ctrlo{-1} & \ctrlo{-1} &\qw & \rstick{\sigma_x,\sigma_y}\\
        &{\hspace{2em} \rho_1 \hspace{0em}}&&&&&  \rho_2 &&{\hspace{1.7em}  \hspace{2em}}&&&&&\rho_3& }\\
        \vspace{0.3cm}      
    \caption{Quantum circuit for continuous-variable selective quantum process tomography.}
    \label{circuitCVproceso}
    \end{figure*}

   The translation operator $T(x)$ acts on a position eigenstate $\ket{y}$ as   
    \begin{equation}
        T(x)\ket{y}=\ket{x+y}.
        \label{accionT}
    \end{equation}

    The action of the squeezing operator $S(r)$, with $r$ a real parameter, on a position eigenstate $\ket{y}$ gives a state proportional to the position eigenstate $\ket{e^{-r}y}$ \cite{kok2010introduction}. Since $S$ is unitary,
    \begin{equation}
        S(r)\ket{y}=e^{-r/2}\ket{e^{-r}y}.
        \label{accionS}
    \end{equation}

    The input state to the circuit is $\rho_0=\ket{\psi}\bra{\psi}\otimes\ket{0}\bra{0}$. 
    The pure state $\ket{\psi}$ can be expressed in the position representation as
\begin{equation}
 \ket{\psi}=\int\limits_{-\infty}^{+\infty}\psi(x)\ket{x}\,dx,
 \label{estadopsi}
 \end{equation}
 where $\psi(x)$ is the wave function of the state. There is some freedom in choosing this input state, provided it satisfies certain properties that will be mentioned later.

  A Hadamard gate acts on the control qubit, yielding the combined state:
     \begin{equation}
         \rho_1=\frac{1}{2}\ket{\psi}\bra{\psi}\otimes\left(\ket{0}\bra{0}+\ket{0}\bra{1}+\ket{1}\bra{0}+\ket{1}\bra{1}\right).
     \end{equation}

The squeezing and translation operations are applied in a controlled manner. A controlled operation acts on a system only if the control qubit is in the state $\ket{1}$. That is to say, the controlled operation $C\textrm{-}U$ acts on the combined state $\ket{\varphi}\ket{1}$ as $C\textrm{-}U(\ket{\varphi}\ket{1})=(U\ket{\varphi})\ket{1}$ and leaves the state $\ket{\varphi}\ket{0}$ unchanged, while for the anticontrolled operation, $U$  acts on $\ket{\varphi}$ only when the control qubit is $\ket{0}$.

The state $\rho_2$, after the action of controlled squeezing and translation operators, is
\begin{equation}
  \begin{split}
      \rho_2=&\frac{1}{2}\left(T(b)S(r_b) \ket{\psi}\bra{\psi} S^\dagger(r_b)T^\dagger(b)\otimes \ket{0}\bra{0}+\right.\\
      &\left.T(c)S(r_c) \ket{\psi}\bra{\psi} S^\dagger(r_c)T^\dagger(c)\otimes \ket{1}\bra{1}+\right.\\
      &\left.T(c)S(r_c) \ket{\psi}\bra{\psi} S^\dagger(r_b)T^\dagger(b)\otimes \ket{1}\bra{0}+\right.\\
      &\left.T(b)S(r_b) \ket{\psi}\bra{\psi} S^\dagger(r_c)T^\dagger(c)\otimes \ket{0}\bra{1}\right).
   \end{split}
 \end{equation}

% \begin{equation}
    %\begin{split}
     %   \rho_3=&\frac{1}{2}\int \mathcal{E}_{\substack{x,y\\w,z}}\left(\ket{x}\bra{y}T(b)S(r_b) \ket{\psi}\bra{\psi} S^\dagger(r_b)T^\dagger(b)\ket{w}\bra{z}\otimes\right.\\
      %  &\left.\ket{0}\bra{0}+ \ket{x}\bra{y}T(c)S(r_c) \ket{\psi}\bra{\psi} S^\dagger(r_c)T^\dagger(c)\ket{w}\bra{z}\otimes\right.\\
       % &\left.\ket{1}\bra{1} +\ket{x}\bra{y} T(c)S(r_c) \ket{\psi}\bra{\psi} S^\dagger(r_b)T^\dagger(b)\ket{w}\bra{z}\otimes\right.\\
        %&\left.\ket{1}\bra{0} + \ket{x}\bra{y}T(b)S(r_b) \ket{\psi}\bra{\psi} S^\dagger(r_c)T^\dagger(c)\ket{w}\bra{z}\otimes \right.\\
        %&\left.\ket{0}\bra{1}\right)\,dx\,dy\,dw\,dz.\\
    %\end{split}
    %\end{equation}
 
    The unknown quantum process $\mathcal{E}$ is then applied to the upper register, followed by
    the controlled operations $T^\dagger (d)$ and $S^\dagger(r_d)$, and the anticontrolled operations $T^\dagger (a)$ and $S^\dagger(r_a)$, yielding the final state $\rho_3$ before the measurement process. This state can be expressed as
 \begin{equation}
    \begin{split}
        \rho_3&=\frac{1}{2}\int \mathcal{E}_{\substack{x,y\\w,z}}\left(S^\dagger(r_a)T^\dagger(a)\ket{x}\bra{y}T(b)S(r_b) \ket{\psi}\right.\\
        &\left.\bra{\psi} S^\dagger(r_b)T^\dagger(b)\ket{w}\bra{z}T(a)S(r_a)\otimes \ket{0}\bra{0}+\right.\\
        &\left. S^\dagger(r_d)T^\dagger(d) \ket{x}\bra{y}T(c)S(r_c) \ket{\psi}\bra{\psi} S^\dagger(r_c)T^\dagger(c)\ket{w}\right.\\
        &\left. \bra{z}T(d)S(r_d)\otimes \ket{1}\bra{1}+S^\dagger(r_d)T^\dagger(d)\ket{x}\bra{y} T(c)S(r_c) \ket{\psi}\right.\\
        &\left.\bra{\psi} S^\dagger(r_b)T^\dagger(b)\ket{w}\bra{z}T(a)S(r_a)\otimes \ket{1}\bra{0}+\right.\\
        &\left. S^\dagger(r_a)T^\dagger(a)\ket{x}\bra{y}T(b)S(r_b) \ket{\psi}\bra{\psi}S^\dagger(r_c)T^\dagger(c)\ket{w}\right.\\
        &\left. \bra{z}T(d)S(r_d)\otimes \ket{0}\bra{1}\Big)\right.\,dx\,dy\,dw\,dz.
    \end{split}
    \end{equation}

    Taking into account that, for continuous-variable systems, a projection measurement cannot be carried out with infinite precision, we define the projection operator around the origin as \cite{pati}
	\begin{equation}
        \Pi_{\delta}=\int\limits_{-\delta/2}^{\delta/2}\ket{\xi}\bra{\xi}\,d\xi,
        \label{proyector}
	\end{equation}
    where the  parameter $\delta$ is set by the measurement device.

   The measured average value of the operator $\Pi_\delta\otimes\sigma_x$, with the projection operator $\Pi_\delta$ defined as in \eqref{proyector}, is given by
\begin{equation}
     \begin{split}
       &\textrm{Tr}\left(\rho_3\int\limits_{-\delta/2}^{\delta/2} \ket{\xi}\bra{\xi} \,d\xi\otimes \sigma_x\right) =e^{-\frac{1}{2}R}\Re\int\limits_{-\infty}^{+\infty}\int\limits_{-\infty}^{+\infty}\int\limits_{-\delta/2}^{\delta/2}\int\mathcal{E}_{\substack{x,y\\w,z}}\times\\
       &\psi(\alpha)\psi^*(\beta)\langle e^{-r_a}\xi+a|x\rangle\langle y|e^{-r_b}\alpha+b\rangle\langle e^{-r_c}\beta+c|w\rangle\times\\
       &\langle z|e^{-r_d}\xi+d\rangle \,dx\,dy\,dw\,dz\, d\xi\, d\alpha \,d\beta=\\
       &e^{-\frac{1}{2}R}\Re\left[\int\limits_{-\infty}^{+\infty}\int\limits_{-\infty}^{+\infty}\int\limits_{-\delta/2}^{\delta/2}\psi(\alpha)\psi^*(\beta)\mathcal{E}_{\substack{\xi e^{-r_a}+a,\alpha e^{-r_b}+b\\\beta e^{-r_c}+c,\xi e^{-r_d}+d}}\,d\xi \,d\alpha \,d\beta\right],
   \label{MeasureRealPart1}
    \end{split}
  \end{equation}
      where $R=r_a+r_b+r_c+r_d$.

The integral in \eqref{MeasureRealPart1} acts as a weighted average, and the input state $\ket{\psi}$ partly determines which elements $\mathcal{E}_{\substack{x,y\\w,z}}$ have the greatest weights in the integration.

For instance, the wave function $\psi(x)$ of $\ket{\psi}$ can be chosen as a real positive-valued function, with non-negligible values only in the interval $[-\Delta/2,+\Delta/2]$ and with a maximum in the center of this interval, $x=0$. These requirements are met, for example, by a squeezed coherent state. The expression \eqref{MeasureRealPart1} can then be approximated as
   \begin{equation}
    \begin{split}
&\textrm{Tr}\left(\rho_3\int\limits_{-\delta/2}^{\delta/2} \ket{\xi}\bra{\xi} \,d\xi\otimes \sigma_x\right) \approx\\
& e^{-\frac{1}{2}R}\Re\left[\int\limits_{-\delta/2}^{\delta/2}\int\limits_{-\Delta/2}^{+\Delta/2}\int\limits_{-\Delta/2}^{+\Delta/2}\psi(\alpha)\psi(\beta)\mathcal{E}_{\substack{\xi e^{-r_a}+a,\alpha e^{-r_b}+b\\\beta e^{-r_c}+c,\xi e^{-r_d}+d}}\,d\alpha \,d\beta \,d\xi \right].
\label{MeasureRealPartaprox}
\end{split}
    \end{equation}

    The translation parameters $a$, $b$, $c$, and $d$, and the squeezing parameters $r_a$, $r_b$, $r_c$, and $r_d$ allow one to select a region in $\mathbb{R}^4$, namely
    
    \begin{equation}
    \begin{split}
        \mathcal{R}_{\substack{ab\\cd}}=&[a-\Delta_a/2,a+\Delta_a/2]\times[b-\Delta_b/2,b+\Delta_b/2]\times\\
        &[c-\Delta_c/2,c+\Delta_c/2]\times[d-\Delta_d/2,d+\Delta_d/2],
        \label{intervals}
    \end{split}
    \end{equation}
    where $\Delta_a=e^{-r_a}\delta$, $\Delta_b=e^{-r_b}\Delta$, $\Delta_c=e^{-r_c}\Delta$, and $\Delta_d=e^{-r_d}\delta$.
    The integration in \eqref{MeasureRealPartaprox} is done over a subregion of $\mathcal{R}_{\substack{ab\\cd}}$, which includes the process element $\mathcal{E}_{\substack{a,b\\c,d}}$ we want to estimate.
    Assuming the process elements  $\mathcal{E}_{\substack{x,y\\w,z}}$ do not exhibit meaningful fluctuations over this region, $\mathcal{E}_{\substack{a,b\\c,d}}$ can then be estimated as
    
\begin{equation}
               \mathcal{E}_{\substack{a,b\\c,d}}^{est}=\dfrac{1}{A} \int\limits_{-\delta/2}^{\delta/2}\int\limits_{-\infty}^{+\infty}\int\limits_{-\infty}^{+\infty}\psi(\alpha)\psi(\beta)\mathcal{E}_{\substack{\xi e^{-r_a}+a,\alpha e^{-r_b}+b\\\beta e^{-r_c}+c,\xi e^{-r_d}+d}}\,d\alpha \,d\beta \,d\xi ,
               \label{estimateeq}
             \end{equation}
    where $A=\delta\left( \int\psi(x) dx\right)^{2}$.

    Hence, an estimate for the real part of $\mathcal{E}_{\substack{a,b\\c,d}}$ can be directly derived from our measurement.
    
    Analogously, by measuring the Pauli operator $\sigma_y$ instead of $\sigma_x$, we can estimate the imaginary part of $\mathcal{E}_{\substack{a,b\\c,d}}$, given that
   \begin{equation}
    \begin{split}
        &\textrm{Tr}\left(\rho_3\int\limits_{-\delta/2}^{\delta/2} \ket{\xi}\bra{\xi} \,d\xi\otimes \sigma_y\right) \approx \\
        & e^{-\frac{1}{2}R}\Im\left[\int\limits_{-\delta/2}^{\delta/2}\int\limits_{-\Delta/2}^{+\Delta/2}\int\limits_{-\Delta/2}^{+\Delta/2}\psi(\alpha)\psi(\beta)\mathcal{E}_{\substack{\xi e^{-r_a}+a,\alpha e^{-r_b}+b\\\beta e^{-r_c}+c,\xi e^{-r_d}+d}} \,d\alpha \,d\beta \,d\xi\right].
    \end{split}
    \label{MeasureImPart1}
    \end{equation}

    The estimated values for the real and imaginary parts of $\mathcal{E}_{\substack{a,b\\c,d}}$ are then weighted averages of $\mathcal{E}_{\substack{x,y\\w,z}}$ over a subregion of $\mathcal{R}_{\substack{ab\\cd}}$ that contains $(a,b,c,d)$.

   Taking into account that controlled operations can be more challenging to implement than non-controlled operations, we also present in Fig. \ref{circuitCVprocesosimp} a simplified quantum circuit that uses fewer controlled operations while still obtaining the same results (Eqs. \eqref{MeasureRealPartaprox} and \eqref{MeasureImPart1}).

It is also noteworthy that, for optical systems, homodyne detection is the most common method for sampling the $x$ quadrature. In this case, we can just as easily select any interval $[x-\delta/2, x+\delta/2]$ as the interval $[-\delta/2,+\delta/2]$ in \eqref{proyector}, and the detector's resolution $\delta$ can be controlled by adjusting the amplitude of the local oscillator. This allows us to eliminate one squeezing operator and one translation operator; however, we still need to independently control the remaining three squeezing and translation parameters.
That said, homodyne detection may not always be the most suitable approach. It is important to highlight that our protocol is also applicable to processes acting on non-optical systems.

\begin{figure*}
 \centering
        \large
        \hspace{-0.5cm}
        \Qcircuit @C=0.4em @R=2.3em {
        \lstick{\ket{\psi}} & \qw & \gate{S(r_c)} & \gate{S(r_b-r_c)} & \gate{T(c)}  & \gate{T(b-c)} & \qw  &  \gate{\mathcal{E}} & \qw  &\gate{T^{\dagger}(d)} & \gate{T^{\dagger}(a-d)} &\gate{S^{\dagger}(r_d)} &  \gate{S^{\dagger}(r_a-r_d)} &\qw  & \rstick{\Pi_{\delta}} \\
        \lstick{\ket{0}} & \gate{H} & \qw & \ctrlo{-1} & \qw & \ctrlo{-1} & \qw & \qw& \qw &  \qw & \ctrlo{-1}  & \qw & \ctrlo{-1} &\qw & \rstick{\sigma_x,\sigma_y}}\\
        \vspace{0.3cm}      
    \caption{Simplified quantum circuit for continuous-variable selective quantum process tomography.}
    \label{circuitCVprocesosimp}
    \end{figure*}
    
    Since it is not possible to know whether the process exhibits meaningful fluctuations in the region \eqref{intervals}, it is necessary to establish a criterion for selecting $\mathcal{R}_{\substack{ab\\cd}}$ in order to find a region where fluctuations are negligible.

Different proposals can be implemented to select the aforementioned region  over which the process element is estimated. One possible approach is detailed as follows.

\begin{enumerate}
    \item We start by setting the squeezing parameters $r_a=r_b=r_c=r_d=0$, thus obtaining an initial region $\mathcal{R}_{\substack{ab\\cd}}^0$:

\begin{equation}
    \begin{split}
        \mathcal{R}_{\substack{ab\\cd}}^0=&[a-\delta/2,a+\delta/2]\times[b-\Delta/2,b+\Delta/2]\times\\
        &[c-\Delta/2,c+\Delta/2]\times[d-\delta/2,d+\delta/2],
        \label{intervals2}
    \end{split}
    \end{equation}

\item Each of the four intervals in \eqref{intervals2} can then be divided into three subintervals of equal length. Consequently, a set of subregions is obtained, one of which is centered on $(a,b,c,d)$. 
An estimate $\mathcal{E}_j^{est}$ of the process element in the center of each subregion $j$, or of a random subset of subregions, can be determined by means of the algorithm presented above. 

\item Fixing a weight $\varepsilon$, we ask for the following condition to be verified:
\begin{equation}
        \textrm{max}_{j}\dfrac{\left|\mathcal{E}_j^{est}-\mathcal{E}_{\substack{a,b\\c,d}}^{est}\right|}{\left|\mathcal{E}_{\substack{a,b\\c,d}}^{est}\right|}\leq \varepsilon,
        \label{condition}
    \end{equation}
    where $\mathcal{E}_{\substack{a,b\\c,d}}^{est}$ is the estimated value for the central region containing $(a,b,c,d)$. If this condition is not met, then the central region is selected as the new initial region, and the procedure is repeated until the aimed condition \eqref{condition} is satisfied, and the region $\mathcal{R}_{\substack{ab\\cd}}$ is defined. 
\end{enumerate}

    By following this procedure, we consider that the fluctuations of $\mathcal{E}_{\substack{x,y\\w,z}}$ over the selected region $\mathcal{R}_{\substack{ab\\cd}}$ are negligible when all the regions $j$ weight the same within an error $\varepsilon$.

     Once the region  $\mathcal{R}_{\substack{ab\\cd}}$ is determined, in order to determine the number of experimental repetitions required to obtain the resulting estimation $\mathcal{E}_{\substack{a,b\\c,d}}^{est}$, we have to take into consideration that each experimental run gives one of three possible outcomes: a click of the detector in the upper register, and a $\pm 1$ result for the measurement of $\sigma_x$ ($\sigma_y$). These results can be labeled as $\pm 1$, or no click of the detector, labeled as 0. Each of these results is detected at random with its corresponding probability.
 
  Therefore, by means of a Chernoff bound \cite{chernoff}, the number of experimental runs $M$ required to determine $\mathcal{E}_{\substack{a,b\\c,d}}^{est}$ having an error equal to or greater than $\epsilon$ with a probability $p$ or less is bounded by
  \begin{equation}
M\geq \dfrac{2\textrm{ln}(\frac{2}{p})}{\epsilon^2A^{2}e^{-R}}.
 \end{equation}
 Thus, the number of experimental repetitions scales polynomially with the uncertainty $\epsilon$.

 \section{III. Numerical simulations}
\label{section3}

 In this section, we numerically simulate the application of our protocol to estimate different continuous-variable quantum processes, described as in \eqref{processdef1}, over a region. 
 
 We firstly take as an illustrative example the Fourier transform process $\mathcal{F}$, which is the continuous-variable analog of the Hadamard gate \cite{braunstein2005quantum}. This operation acts on a position eigenstate as

\begin{equation}
 \mathcal{F}\ket{x}=\dfrac{1}{\sqrt{2\pi}}\int\limits_{-\infty}^{+\infty}e^{ixy}\ket{y}dy=\ket{p_x},
\end{equation}
where $\ket{p_x}$ is the $x$-momentum eigenstate. 

The action of the Fourier transform on a density operator $\rho$ can then be written as

\begin{equation}
\mathcal{F}(\rho)=\mathcal{F}\rho\mathcal{F}^\dagger=\dfrac{1}{2\pi}\int e^{i(xy-wz)}\ket{x}\bra{y}\rho\ket{w}\bra{z}dxdydxdz.
\end{equation}

Therefore, this process can be completely described by
\begin{equation}
\mathcal{E}_{\substack{x,y\\w,z}}^{\mathcal{F}}=\dfrac{1}{2\pi}e^{i(xy-wz)}.
\label{cubicop}
\end{equation}

The importance of the Fourier transform operation for the development of quantum computation with continuous variables is noteworthy. This process alongside other operations, can form a universal gate set for continuous-variable quantum computation. For instance, the set conformed by the displacement, the squeezing, the beam-splitter, and the Fourier transform operators forms a universal set for Gaussian operations \cite{lloyd1999quantum,hillmann2020universal}. Addition of the non-Gaussian cubic phase gate, $e^{i\gamma\hat{x}^3}$, where $\hat{x}$ represents the position operator and $\gamma$ is a real parameter, allows for universal quantum computation. Taking this into account, we also numerically simulate the reconstruction over a region of a non-Gaussian operation, $\mathcal{G(\gamma)}$, defined as

\begin{equation}
    \mathcal{G}(\gamma)=e^{i\gamma \hat{x}^3}\mathcal{F}.
\end{equation}

The action of this operator on a position eigenstate can be expressed as
\begin{equation}
    \mathcal{G}(\gamma)\ket{x}=\dfrac{1}{\sqrt{2\pi}}\int\limits_{-\infty}^{\infty}e^{i\gamma y^3}e^{ixy}\ket{y}dy.
\end{equation}

Therefore, this process can be completely described by

\begin{equation}
   \mathcal{E}_{\substack{x,y\\w,z}}^{\mathcal{G}}=\dfrac{1}{2\pi}e^{i\gamma(x^3-z^3)}e^{i(xy-wz)}.
\end{equation}

Finally, we numerically simulate the reconstruction over a region of a quantum process, $\mathcal{U}(r,\kappa)$, defined as
\begin{equation}
    \mathcal{U}(r,\kappa)=\mathcal{F}S(r)P(\kappa),
    \label{gaussop}
\end{equation}
where $P(\kappa)=e^{i\kappa\hat{x}^2}$ is the quadratic phase gate, with $\kappa$  a real parameter. It is worth highlighting that any single-mode Gaussian unitary can be decomposed into a finite sequence of operations $\mathcal{U}(r,\kappa)$ (up to displacements) \cite{ukai2010,alexander2014}.

Taking $r$ as a real parameter, the action of $\mathcal{U}(r,\kappa)$ on a position eigenstate is given by
\begin{equation}
    \mathcal{U}(r,\kappa)\ket{x}=\dfrac{1}{\sqrt{2\pi}}\int\limits_{-\infty}^{\infty}e^{-\kappa x^2-r/2}e^{ie^{-r}xy}\ket{y}dy.
\end{equation}

Therefore, this process can be completely described by
\begin{equation}
   \mathcal{E}_{\substack{x,y\\w,z}}^{\mathcal{U}}=\dfrac{1}{2\pi}e^{i\kappa(y^2-w^2)-r}e^{ie^{-r}(xy-wz)}.
\end{equation}

We set our initial state $\ket{\psi}$ as a coherent squeezed state, whose wave function is given by
\begin{equation}
    \psi(x)\propto e^{-Cx^2},
\end{equation}
where $C$ is a constant such that $\psi(x)<0.05$ for $|x|>\Delta/2$, taking $\Delta=0.1$. The state $\ket{\psi}$ is normalized.

We start by restricting the region over which we want a description of our process. In this case, we numerically simulate the estimation of $\mathcal{E}_{\substack{x,y\\w,z}}$, for $x,y,w,z\in [0,3]$.  A mesh over this region is defined, with equidistant points where the elements $\mathcal{E}_{\substack{x,y\\w,z}}$ will be estimated. For each point $(a,b,c,d)$ of the mesh, we proceed as follows.
\begin{enumerate}
    \item By setting the squeezing parameters as zero, the initial region $\mathcal{R}_{\substack{ab\\cd}}^0$ \eqref{intervals2} centered around the mesh point $(a,b,c,d)$ is selected.
    \item  Partitioning each of the four intervals in  \eqref{intervals2} in three subintervals of equal length, the initial region $\mathcal{R}_{\substack{ab\\cd}}^0$ is thus divided into subregions of equal volume, one of which is centered on the mesh point $(a,b,c,d)$. 
    \item An estimated value of $\mathcal{E}_{\substack{x,y\\w,z}}^{est}$ is determined for the central point of each subregion, by means of \eqref{estimateeq}.
    \item Condition \eqref{condition} is tested. If the condition is not verified, the central region containing the mesh point $(a,b,c,d)$ is taken as the new initial region and the above steps are repeated. This procedure is repeated until condition \eqref{condition} is fulfilled.
    
\end{enumerate}

\begin{figure*}
\subfloat[ \label{figRe3D}]{%
  \includegraphics[width=0.49\linewidth]{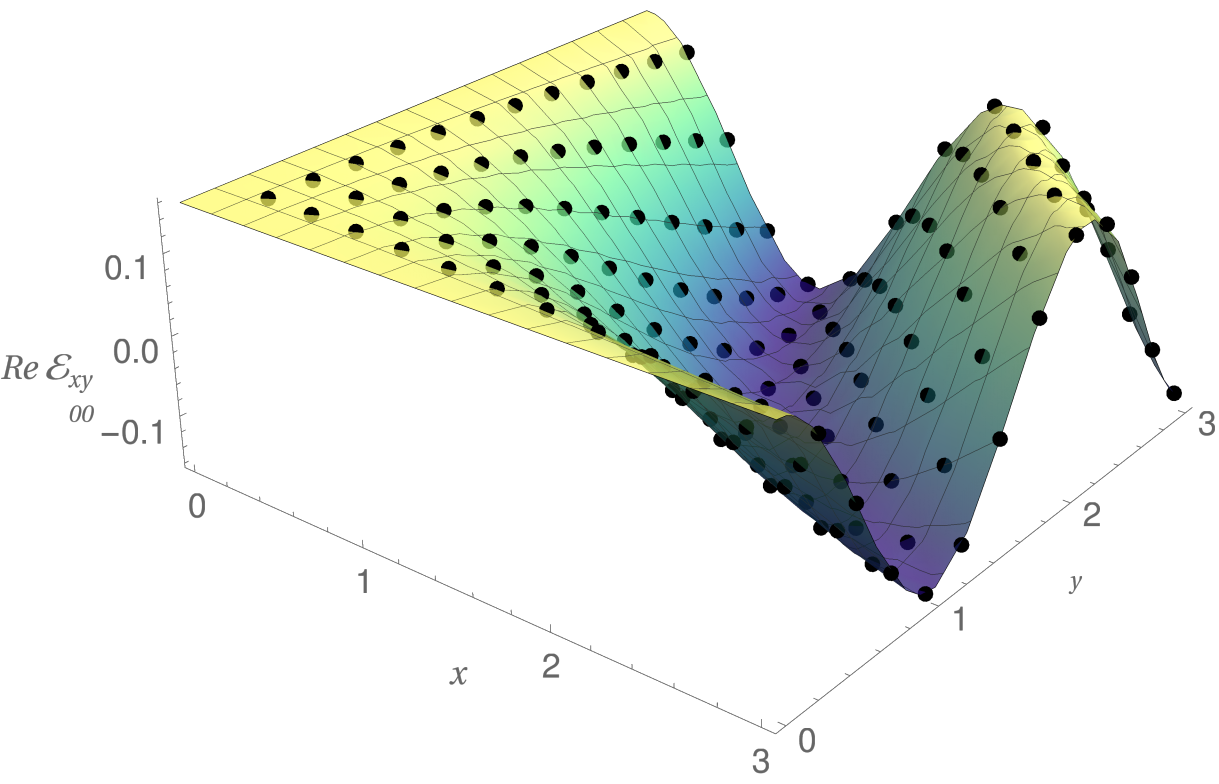}  
}\hfill
\subfloat[ \label{figIm3D}]{
\includegraphics[width=0.49\linewidth]{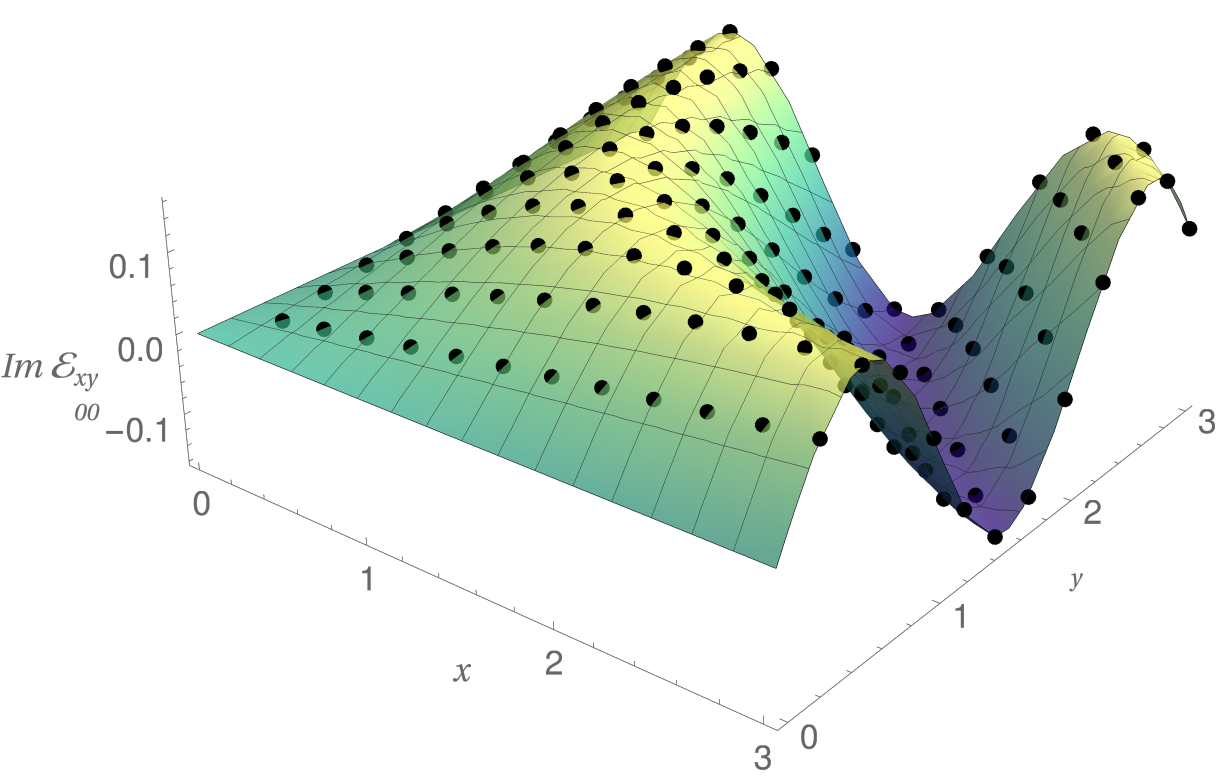}  
}
\caption{Theoretical form and estimated values of the real (a) and imaginary (b) parts of $\mathcal{E}_{\substack{x,y\\0,0}}$ for the Fourier transform, $\mathcal{F}$, with $x,y\in [0,3]$. The estimated $\mathcal{E}_{\substack{x,y\\0,0}}^{est}$ for each  mesh point are shown as black points.
The error $\epsilon$ in condition \eqref{condition} was set as $\epsilon=0.05$. The detector precision $\delta$ and the parameter $\Delta$ were set equal to $0.1$.}

\label{fig3D}
\end{figure*}

\begin{figure*}
\subfloat[ \label{figRe3D}]{%
  \includegraphics[width=0.49\linewidth]{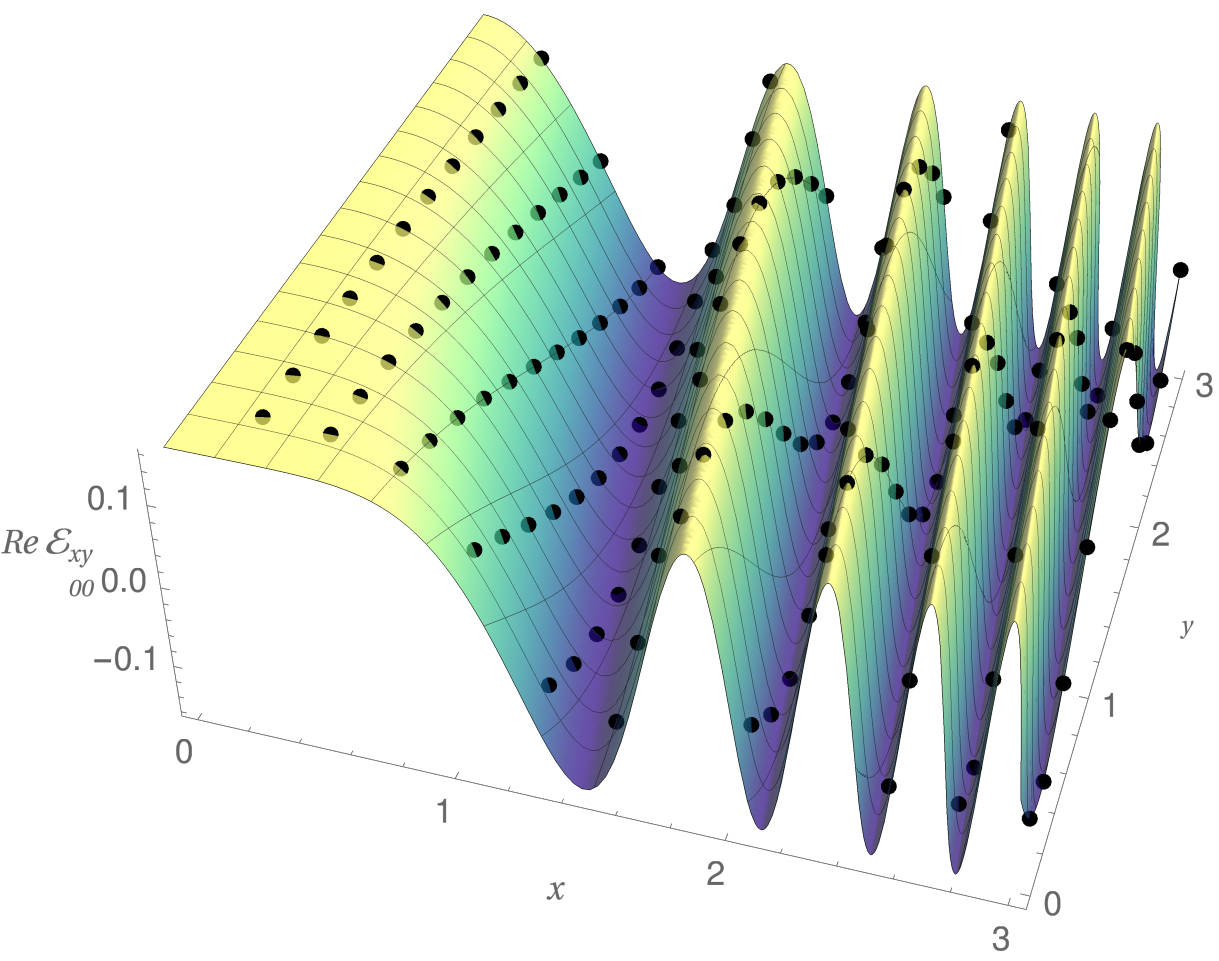}  
}\hfill
\subfloat[ \label{figIm3D}]{
\includegraphics[width=0.49\linewidth]{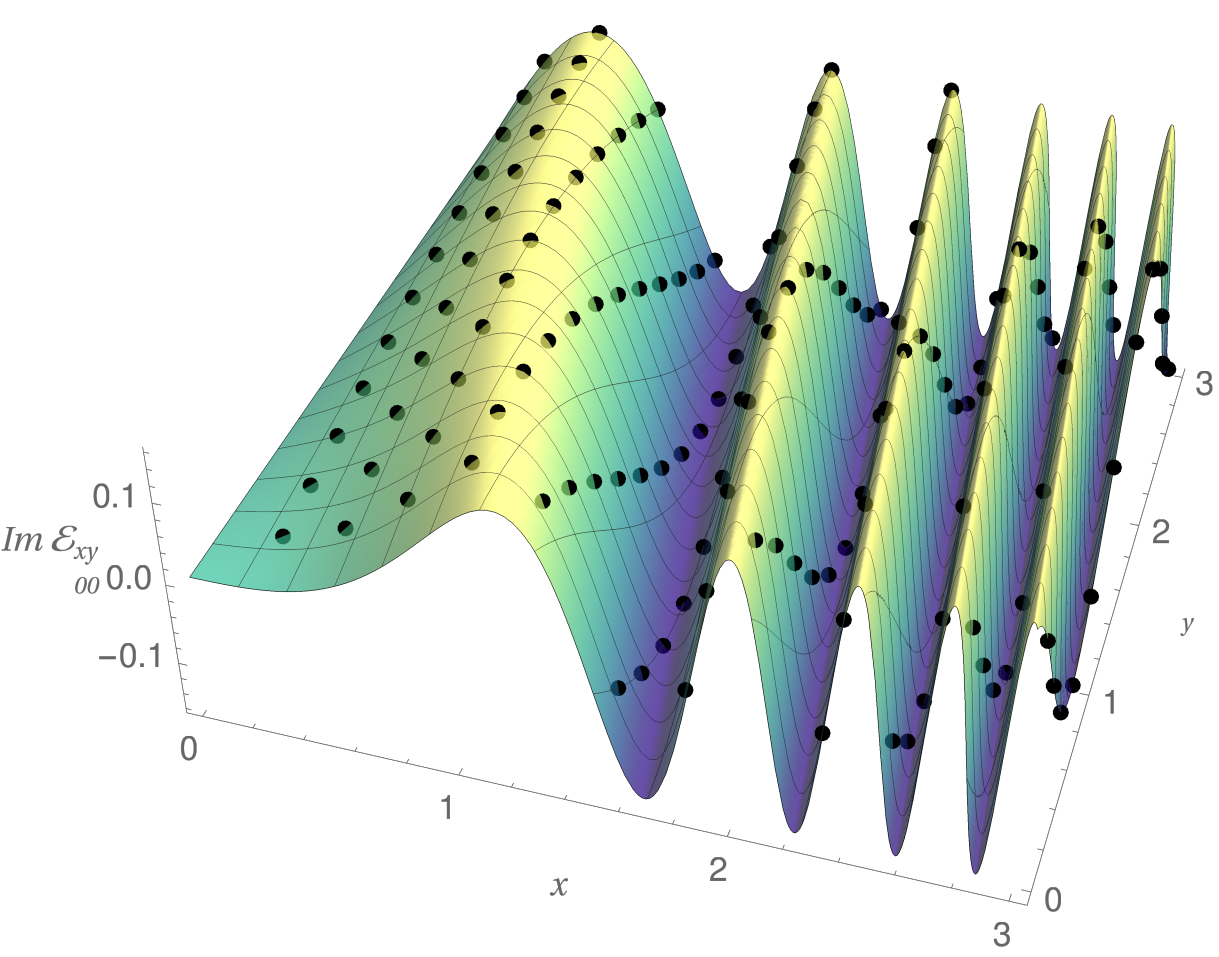}  
}
\caption{Theoretical form and estimated values of the real (a) and imaginary (b) parts of $\mathcal{E}_{\substack{x,y\\0,0}}$ for the process $\mathcal{G}(\gamma=1)$ \ref{cubicop}, with $x,y\in [0,3]$. The estimated $\mathcal{E}_{\substack{x,y\\0,0}}^{est}$ for each  mesh point are shown as black points.
The error $\epsilon$ in condition \eqref{condition} was set as $\epsilon=0.05$. The detector precision $\delta$ and the parameter $\Delta$ were set equal to $0.1$.}

\label{fig3Dcubic}
\end{figure*}

\begin{figure*}
\subfloat[ \label{figRe3D}]{%
  \includegraphics[width=0.49\linewidth]{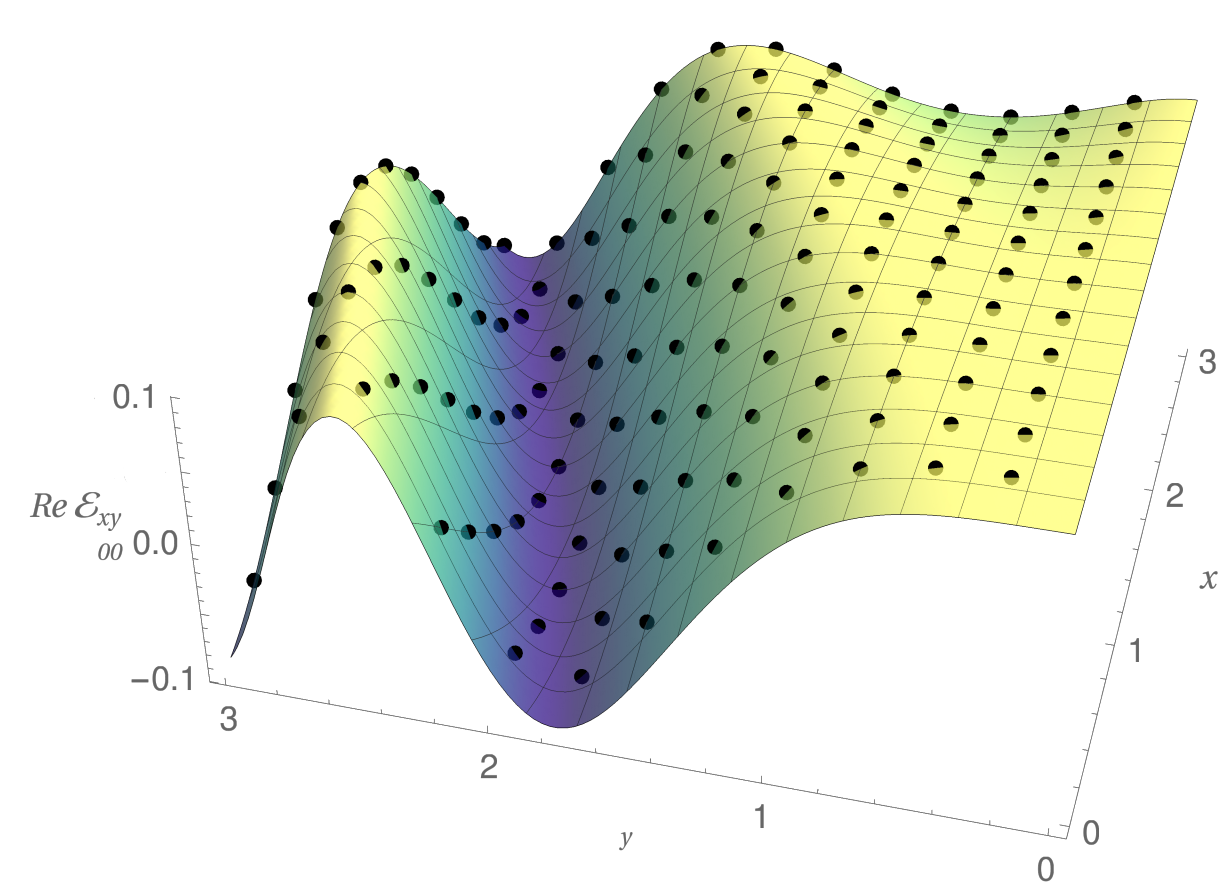}  
}\hfill
\subfloat[ \label{figIm3D}]{
\includegraphics[width=0.49\linewidth]{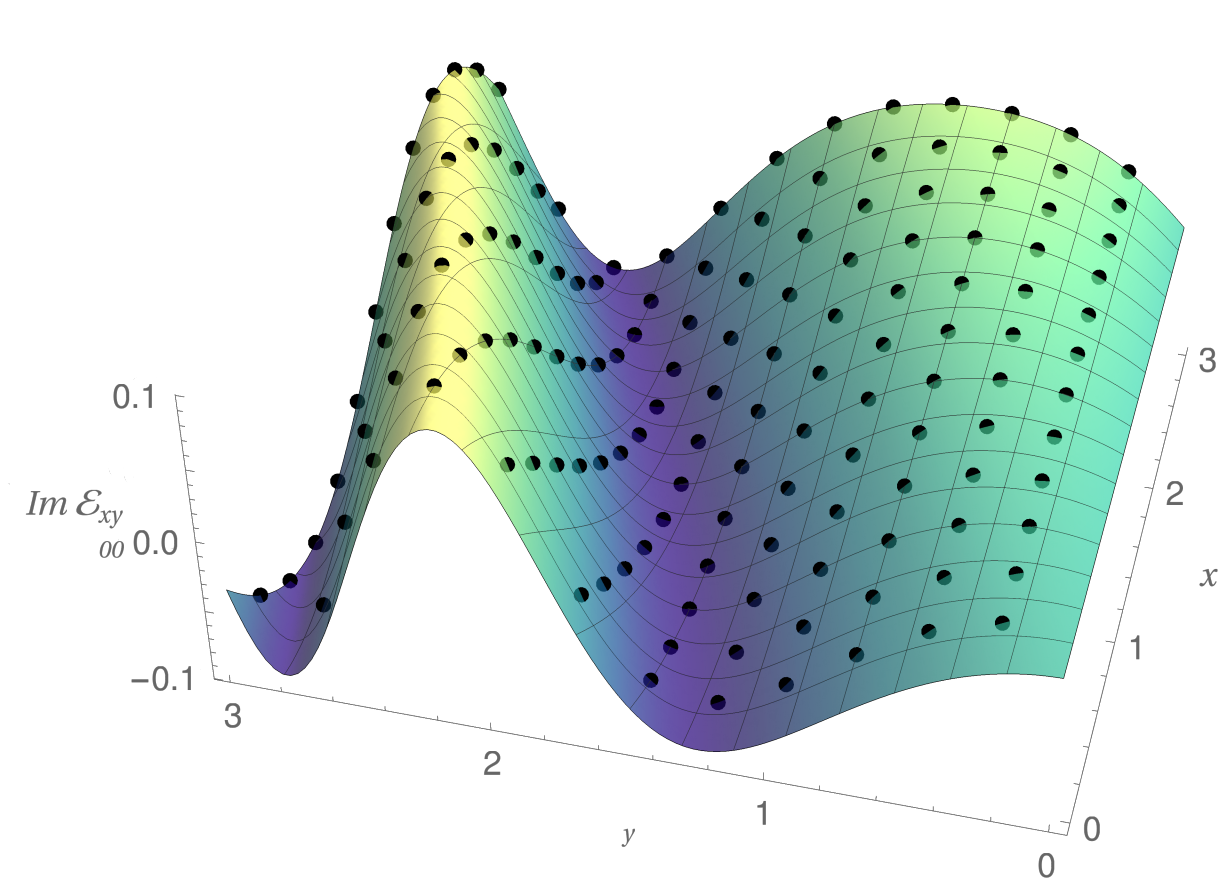}  
}
\caption{Theoretical form and estimated values of the real (a) and imaginary (b) parts of $\mathcal{E}_{\substack{x,y\\0,0}}$ for the process $\mathcal{U}(r=0.5,\kappa=-1)$ \ref{gaussop}, with $x,y\in [0,3]$. The estimated $\mathcal{E}_{\substack{x,y\\0,0}}^{est}$ for each  mesh point are shown as black points.
The error $\epsilon$ in condition \eqref{condition} was set as $\epsilon=0.05$. The detector precision $\delta$ and the parameter $\Delta$ were set equal to $0.1$.}

\label{fig3Dgaussian}
\end{figure*}

The estimated and theoretical real and imaginary parts of $\mathcal{E}_{\substack{x,y\\0,0}}$, with $x,y\in [0,3]$, for $\mathcal{F}$, $\mathcal{G}(\gamma=1)$ and $\mathcal{U}(r=0.5,\kappa=-1)$  are presented in Figs. \ref{fig3D}, \ref{fig3Dcubic} and \ref{fig3Dgaussian},respectively. 

For the three studied cases, the maximum relative error, $e_{max}$, between the simulated estimates and the theoretical values is of the order
\begin{equation}
e_{max}=\textrm{max}\dfrac{\left|\mathcal{E}_{\substack{x,y\\w,z}}^{est}-\mathcal{E}_{\substack{x,y\\w,z}}\right|}{\left|\mathcal{E}_{\substack{x,y\\w,z}}\right|}\sim 10^{-3}.
 \end{equation}

In order to quantify how much the simulated estimation over the studied region differs from the theoretical results, we found bounds for the fidelities between the corresponding isomorphic states.  

The Choi–Jamiołkowski isomorphism \cite{choi1975completely,jamiolkowski1972linear} states that there is a one-to-one correspondence between quantum states and quantum channels in finite dimension. This duality has been extended so it is valid in infinite dimensions. In \cite{kiukas2017continuous} it has been proved that there is a one-to-one correspondence between the family of bipartite states $\rho_{\mathcal{E}}\in \mathcal{H}_A\otimes\mathcal{H}_B$ for which $\textrm{Tr}_A(\rho)=\mu$ and quantum channels $\mathcal{E}$ of dimension $d$, where $d$ can be infinite, such that 
  \begin{equation}
     \rho_{\mathcal{E}}=(\mathcal{E}\otimes \textrm{Id})\ket{\Omega_\mu}\bra{\Omega_\mu},
     \label{estadoiso}
 \end{equation}
  where 
 \begin{equation}
   \ket{\Omega_\mu}=\sum\limits_{n=0}^{d}\sqrt{s_n}\ket{n}\ket{n}  
   \label{purification}
 \end{equation}
 is the purification of $\mu=\sum\limits_ns_n\ket{n}\bra{n}$, expressed in the Fock basis $\{\ket{n}\}$.

Particularly, for infinite dimensional processes the state $\ket{\Omega_\mu}$ can be taken as the two-mode squeezed vacuum:
 \begin{equation}
     \ket{\Omega_\mu}=\sqrt{1-\lambda^2}\sum\limits_{n=0}^{\infty}\lambda^n\ket{n}\ket{n},
     \label{estadoiso2}
 \end{equation}
 where $\lambda=\textrm{tanh}r$, with $r$ the state squeezing parameter.
 This description of quantum processes contains all the information of the channel, and, in that sense, it is equivalent to the description \eqref{processdef1}.
 
To relate the matrix elements $\bra{x}\bra{x'}\rho_{\mathcal{E}}\ket{y'}\ket{y'}$ with the process elements $\mathcal{E}_{\substack{a,b\\c,d}}$ it can be shown that
\begin{equation}
\begin{split}
&\rho_{\mathcal{E}}=\\
&(1-\lambda^2)\sum_{n,m}\lambda^n\lambda^m\int \mathcal{E}_{\substack{x,y\\w,z}}\ket{x}\langle y|n\rangle\langle m|w\rangle\bra{z} dx dy dw dz\otimes \ket{n}\bra{m}\\
    &=(1-\lambda^2)\sum_{n,m} \mathcal{E}_{\substack{x,y\\w,z}}\psi_n(y)\psi_m(w)\ket{x}\bra{z} dx dy dw dz \otimes \ket{n}\bra{m},
\end{split}
\end{equation}
where $\psi_n(x)=\bra{n} x\rangle=\dfrac{H_n(x)}{\pi^{1/4}\sqrt{2^nn!}}e^{-x^2/2}$ is the wave function of the Fock state $\ket{n}$, with $H_n(x)$ the Hermite polynomials.

Then a matrix element $\bra{a}\bra{b}\rho_{\mathcal{E}}\ket{d}\ket{c}$ can be expressed as
\begin{equation}
\begin{split}
&\bra{a}\bra{b}\rho_{\mathcal{E}}\ket{d}\ket{c}=(1-\lambda^2)\int\mathcal{E}_{\substack{a,y\\w,d}}\left(\sum_n \lambda^n\psi_n(y)\psi_n(b)\right)\times\\
&\left(\sum_m \lambda^m \psi_m(w)\psi_m(c)\right) dy dw=\\
&(1-\lambda^2)\int\mathcal{E}_{\substack{a,y\\w,d}}f_\lambda(y,b)f_\lambda(w,c)dydw.
\label{relacioniso1}
\end{split}
\end{equation}
Here we have defined the function
\begin{equation}
f_\lambda(x,y)=
\sum\limits_{n=0}^\infty \lambda^n\psi_n(x)\psi_n(y),
\end{equation}
with $-1<\lambda<1$. This function can be expressed as \cite{mehler,bateman}

\begin{equation}
\begin{split}
&f_\lambda(x,y)=\\
&\dfrac{1}{\sqrt{\pi(1-\lambda^2)}}e^{\left( -\dfrac{(1-\lambda)(x+y)^2}{4(1+\lambda)}-\dfrac{(1+\lambda)(x-y)^2}{4(1-\lambda)}\right)}.
\end{split}
\end{equation}

 The fidelity between two quantum states $\rho_1$ and $ \rho_2$ can be bounded by their trace distance  as \cite{fuchs1999cryptographic}
\begin{equation}
    1-\sqrt{F(\rho_1,\rho_2)}\leq T(\rho_1,\rho_2),
    \label{boundfidelity}
\end{equation}
where $T(\rho_1,\rho_2)$ represents the trace distance between $\rho_1$ and $\rho_2$. Specifically,
\begin{equation}
    T(\rho_1,\rho_2)=\dfrac{1}{2}||\rho_1-\rho_2||_1=\frac{1}{2}\textrm{Tr}|\rho_1-\rho_2|.
\end{equation}

A bound for the trace distance between the theoretical isomorphic state and the simulated estimation of the isomorphic state was found for each case by taking the maximum error between the theoretical process elements $\mathcal{E}_{\substack{x,y\\w,z}}$ and the simulated estimations as the maximum error obtained for the studied region and applying expression \eqref{relacioniso1}. The error between the matrix elements of the theoretical isomorphic state and those of the estimated isomorphic state  was considered only where the isomorphic state takes non-negligible values. Choosing $\lambda=0.8$ in \eqref{estadoiso2} to construct the isomorphic state \eqref{estadoiso} and considering \eqref{relacioniso1} and \eqref{boundfidelity}, the lower bounds, $F_{min}$, for the fidelities between the theoretical isomorphic states $\rho_{\mathcal{E}}$ and its simulated estimations $\rho_{\mathcal{E}}^{est}$ are shown in Table \ref{lowerboundF}.

\begin{table}[ht]
\centering
\setlength{\tabcolsep}{8pt}
\begin{tabular}[t]{lcc}
\hline
\hline
& $\mathcal{E}_{\substack{x,y\\w,z}}$\vspace{0.1cm} & $F_{min}$\\
\hline
$ \mathcal{F}$& $\dfrac{1}{2\pi}e^{i(xy-wz)}$ & 0.98\\ [2ex]
$\mathcal{G}(\gamma)$& $\dfrac{1}{2\pi}e^{i\gamma (x^3-z^3)}e^{i(xy-wz)}$ &0.96\\ [2ex]
$\mathcal{U}(r,\kappa)$& $\dfrac{1}{2\pi}e^{i\kappa(w^2-y^2)-r}e^{i(e^{-r}(xy-wz))}$  & 0.98\\ [2ex]
\hline
\hline
\end{tabular}
\caption{Lower bounds ($F_{min}$) for the fidelities between the theoretical isomorphic states, $\rho_\mathcal{E}$, and its simulated estimations, $\rho_{\mathcal{E}}^{est}$, $F(\rho_\mathcal{E},\rho_{\mathcal{E}}^{est})\geq F_{min}$.}
\label{lowerboundF}
\end{table}

	\begin{figure*}
\subfloat[ \label{figRe}]{%
  \includegraphics[width=0.49\linewidth]{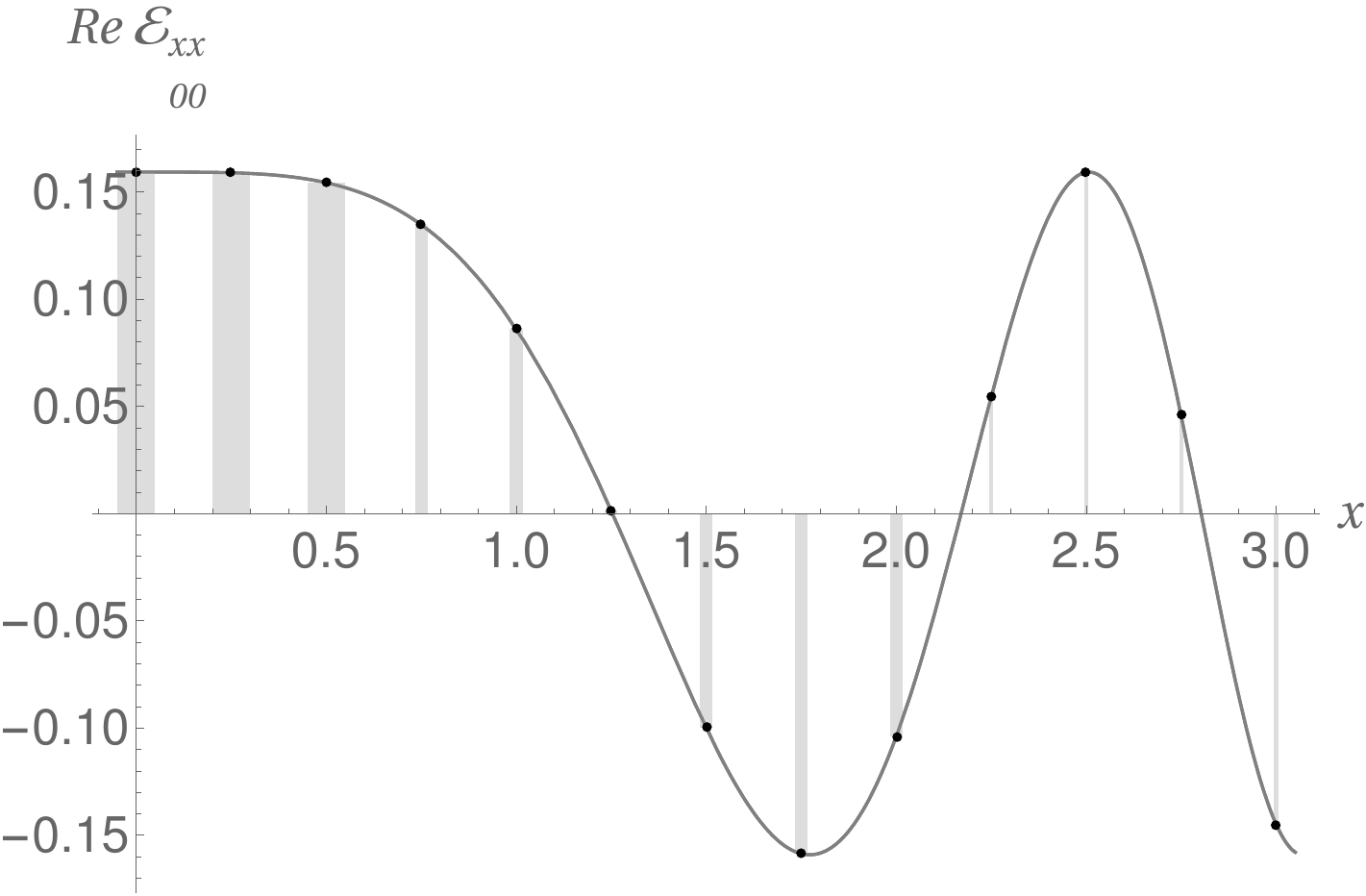}  
}\hfill
\subfloat[ \label{figIm}]{
\includegraphics[width=0.49\linewidth]{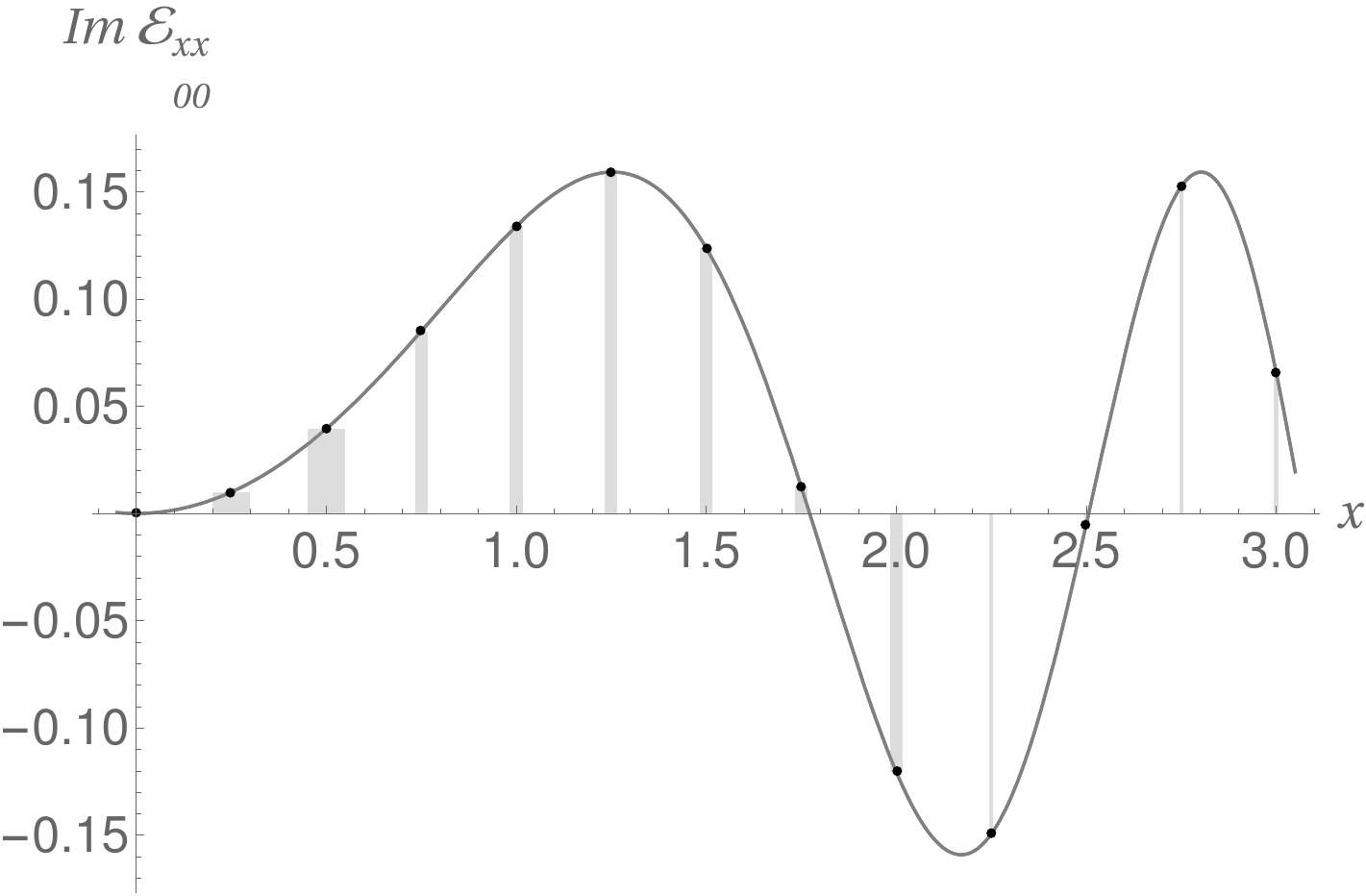}  
}
\caption{Theoretical form and estimated values of the real (a) and imaginary (b) parts of $\mathcal{E}_{\substack{x,x\\0,0}}$ for the Fourier transform, $\mathcal{F}$, with $x\in [0,3]$. The estimated  $\mathcal{E}_{\substack{x,x\\0,0}}^{est}$ values for each  mesh point are shown as black points. The error $\epsilon$ in condition \eqref{condition} was set as $\epsilon=0.05$. The detector precision $\delta$ and the parameter $\Delta$ were set equal to $0.1$. The gray rectangles represent the size of the region for which condition \eqref{condition}  is satisfied and over which the integration in  \eqref{estimateeq} was carried out.}
\label{figregions}
\end{figure*}

It should be noted that these results are just a measure of the distance between the simulated estimations of the processes and the theoretical processes in the studied regions. For other regions, meaningful fluctuations of $\mathcal{E}_{\substack{x,y\\w,z}}$  could be found, which can lead to greater estimation errors.

In Fig. \ref{figregions} the volume of the region $\mathcal{R}_{\substack{ab\\cd}}$ \eqref{intervals} over which the process element $\mathcal{E}_{\substack{x,x\\0,0}}$, for $x\in [0,3]$, was estimated is shown for the Fourier transform case. As expected, smaller regions of the form \eqref{intervals} need to be defined when $\mathcal{E}_{\substack{x,y\\w,z}}$ exhibits greater fluctuations over smaller regions: if $\mathcal{E}_{\substack{x,y\\w,z}}$ fluctuates significantly over the region \eqref{intervals}, the weighted average \eqref{estimateeq} will likely lead to a greater error in the estimated value of the element $\mathcal{E}_{\substack{a,b\\c,d}}$. Defining smaller regions of the form \eqref{intervals} implies greater squeezing; consequently, meaningful fluctuations of  $\mathcal{E}_{\substack{x,y\\w,z}}$ over small regions entail a limitation for this algorithm. This limitation will also be inevitably found in other protocols for continuous-variable quantum process tomography.

\section{IV. Discussion}
\label{section4}

In this paper we proposed a protocol that allows one to directly estimate from measurements any continuous-variable process element in the position representation. This algorithm is selective and works without requiring the complete reconstruction of the process or of its isomorphic state, nor does it rely on inverse linear transform techniques or statistical inference methods, which are frequently used in QPT and can be computationally expensive. Additionally, it does not make use of a set of probing states, as is commonly employed in the most conventional QPT schemes in infinite dimensions, where quantum state tomography is used in order to reconstruct the output states after the action of the unknown process on a set of coherent states \cite{lobino2008complete,rahimi2011quantum,anis2012maximum,kumar2013experimental,lobino2009memory}. The proposed algorithm works with a single input state, which can be selected from a range of possible states that are currently implementable.

For this protocol to be carried out, an
efficient implementation of the controlled squeezing and
translation gates is necessary. Such an implementation is of the utmost importance for the development of quantum computation with continuous variables \cite{braunstein2005quantum,lloyd1999quantum,menicucci2014fault,gu2009quantum}. Significant progress towards this end has been made in recent years. For instance, Ref. \cite{drechsler2020state} proposes the implementation of a controlled
squeezing gate for the squeezing of trapped ions. It is also
noteworthy that displacement and squeezing gates,
along with other operations, can form a universal gate set for
continuous-variable quantum computation \cite{lloyd1999quantum,hillmann2020universal}.

We showed how this protocol can be used to reconstruct a continuous-variable quantum process on a region with high fidelity. However, cases in which the process elements fluctuate significantly over small regions remain a limitation found as well in other QPT protocols, which inevitably arises when working with continuous variables. 

\section{Acknowledgments}
This work was partially supported by Programa de Desarrollo de Ciencias B\'asicas (PEDECIBA, Uruguay), Comisi\'on Acad\'emica de Posgrado (CAP, UdelaR, Uruguay).

\bibliographystyle{unsrt}
\bibliography{references.bib}

\end{document}